\documentclass[letterpaper]{jpconf}
\usepackage{graphicx}
\newcommand{\xvec}[0]{{\stackrel{\rightharpoonup}{x}}}
\newcommand{\dxvec}[0]{{\Delta \!\! \stackrel{\rightharpoonup}{x}}}
\newcommand{\omg}[0]{{\hat{\Omega}}}
\newcommand{\HpPP}[0]{{H}/({P_{1,i} P_{2,i}})}
\begin{document}
\title{A radiometer for stochastic gravitational waves}

\author{Stefan W. Ballmer}

\address{LIGO Laboratory, Massachusetts Institute of Technology, Cambridge, MA 02139, USA}
\ead{sballmer@ligo.mit.edu}

\begin{abstract}
The LIGO Scientific Collaboration recently reported a new upper limit on an isotropic stochastic background of gravitational waves obtained based on
the data from the 3rd LIGO science Run (S3). Now I present a new method for obtaining directional upper limits
that the LIGO Scientific Collaboration intends to use for future LIGO science runs and that essentially implements a gravitational wave radiometer.
\end{abstract}.

\section{Introduction}
The LIGO Scientific Collaboration
analyzed the data from the 3rd LIGO science run (S3) looking for 
an isotropic background of gravitational waves. In Ref. \cite{s3stoch} it was reported that
a 90\%-confidence Bayesian upper limit on $\Omega_{\rm gw}(f)$ of $8.4 \times 10^{-4}$ was achieved.
$\Omega_{\rm gw}(f)$ is the energy density per logarithmic frequency interval associated with
gravitational waves normalized by the critical energy density $\rho_c$ to close the universe.
However it is possible that the dominant source of stochastic gravitational waves in the LIGO frequency band comes
from an ensemble of astrophysical sources (e.g. \cite{tania,bildsten}). If such an ensemble also turns out to be
dominated by its strongest members the assumption of isotropy is no longer very good. 
Then one should look for anisotropies in the stochastic gravitational wave background.
This was addressed in Ref. \cite{allenottewill,cornish}, but
they focused on getting optimal estimates for each spherical harmonic.
If the stochastic gravitational wave background is indeed dominated by individual sources one can get
a better signal-to-noise ratio by obtaining optimal filters for small patches of the sky.

I present such a directional method that essentially implements a gravitational wave radiometer.
The algorithm has been implemented and will be used to analyze
future LIGO science runs starting with S4.

\section{Search for an isotropic background}
The data from the first 3 LIGO science runs was analyzed with a method described in detail
in Ref. \cite{allenromano,s3stoch,s1stoch}. The data streams from a pair of detectors were cross-correlated
with a cross-correlation kernel $Q$ chosen to be optimal for an assumed strain power spectrum
$H(f)\!=\!H(|f|)$
and angular distribution $P(\omg)\!=\!1$ (isotropic distribution).
Specifically, with $S_1(f)$ and
$S_2(f)$ representing the Fourier transforms of the strain
outputs of two detectors, this cross-correlation is computed in frequency domain segment by segment as:
\begin{equation}\label{e:xcorr}
    Y = \int_{-\infty}^{\infty}df\ \int_{-\infty}^{\infty} df'
        \delta_T(f-f')\,S_1^*(f)\,Q(f')\,S_2(f')
\end{equation}
where $\delta_T$ is a finite-time approximation to the Dirac delta
function. The optimal filter $Q$ has the form:
\begin{equation}\label{e:optfilt}
Q(f) = \lambda \frac{ \gamma_{\rm iso}(f) H(f)}{P_1(f) P_2(f)}
\end{equation}
where $\lambda$ is a normalization factor, $P_1$ and $P_2$ are the
strain noise power spectra of the two detectors, $H(f)$ is
the strain power spectrum of the stochastic background being
searched for ($=S_{\rm gw}(f)$ in Ref. \cite{s3stoch,s1stoch}) and the factor $\gamma_{\rm iso}$ takes into 
account the cancellation of an isotropic omni-directional signal ($P(\omg)\!=\!1$)
at higher frequencies due to the detector separation.
$\gamma_{\rm iso}$ is called the
overlap reduction function \cite{flan} and is given by (the normalization is such
that $\gamma_{\rm iso}(f\!=\!0)\!=\!1$ for aligned and co-located detectors):
\begin{equation}\label{e:gamma_iso}
    \gamma_{\rm iso}(f)=\frac{5}{8 \pi} \sum_A \int_{S^2} d\omg \,\, e^{i 2 \pi f \omg \cdot 
    \frac{\dxvec}{c}}
    \,\, F_1^A(\omg) F_2^A(\omg)
\end{equation}
where $\dxvec=\xvec_2 - \xvec_1$ is the detector separation vector and 
\begin{equation}\label{e:F}
    F_i^A(\omg) = e_{ab}^A(\omg) \frac{1}{2}(\hat{X}_i^a \hat{X}_i^b - \hat{Y}_i^a \hat{Y}_i^b)
\end{equation}
is the response of detector i to a zero frequency, unit amplitude, $A=+,\times$ polarized gravitational wave.

The optimal filter $Q$ is derived assuming that the intrinsic detector
noise is Gaussian and stationary over the measurement time,
uncorrelated between detectors, and uncorrelated with and much
greater in power than the stochastic gravitational wave signal. Under these
assumptions the expected variance, $\sigma^2_Y$, of the
cross-correlation is dominated by the noise in the individual
detectors, whereas the expected value of the cross-correlation $Y$
depends on the stochastic background power spectrum:
\begin{equation}\label{e:sigma}
    \sigma_Y^2 \equiv \langle Y^2 \rangle - \langle Y \rangle^2 \approx
               \frac{T}{4} \left( Q,Q \right)
\: , \,\,\,\,\,\,\,\,\,\,\,\,
    \langle Y \rangle = T \left( Q, \frac{\gamma_{\rm iso} H}{P_1 P_2}\right)
\end{equation}
Here the scalar product $\left(\cdot,\cdot\right)$ is defined as
$\left(A,B\right) = \int_{-\infty}^{\infty} A^*(f) B(f) P_1(f) P_2(f) df$
and $T$ is the duration of the measurement.

To deal with the long-term
non-stationarity of the detector noise the data set from a
given interferometer pair is divided into equal-length intervals,
and the cross-correlation $Y$ and theoretical $\sigma_Y$ are
calculated for each interval, yielding a set $\{Y_I,
\sigma_{Y_I}\}$ of such values, with $I$ labeling the intervals.
The interval length can be chosen such that the detector noise is relatively
stationary over one interval. In Ref. \cite{s3stoch,s1stoch} the interval
length was chosen to be 60~sec. The cross-correlation values
are combined to produce a final cross-correlation estimator,
$Y_{\rm opt}$, that maximizes the signal-to-noise ratio, and has
variance $\sigma_{\rm opt}^2$:
\begin{equation}\label{e:Yopt}
  \begin{array}{cc}
    Y_{\rm opt} = \sum_I \sigma_{Y_I}^{-2} Y_I / \sigma_{\rm
    opt}^{-2}\ ,
    &
\,\,\,\,\,\,\,\,\,\,\,\,
    \sigma_{\rm opt}^{-2} = \sum_I \sigma_{Y_I}^{-2}\ .
  \end{array}
\end{equation}

Since the LIGO Hanford and Livingston sites are separated by 3000km the 
overlap reduction function for this pair has already dropped below 5\% around each
interferometer's sweet spot of 150~Hz.
One unfortunate drawback of this analysis thus is the limited use it makes of the
individual interferometer's most sensitive frequency region.
Moreover, if the dominant gravitational wave background would be
of astrophysical origin the assumption of an isotropic background is not well justified.
If for example the signal is dominated by a few strong sources a directed search can achieve
a better signal-to-noise ratio.

\section{Directional search: a gravitational wave radiometer}
A natural generalization of the method described above can be achieved by finding the optimal
filter for an angular power distribution $P(\omg)$. In this case Eq. \ref{e:sigma}b generalizes to
\begin{equation}\label{e:muPOmega}
    \langle Y \rangle = T \left( Q, \frac{\int_{S^2} d\omg \gamma_{\omg} P(\omg) H}{P_1 P_2}\right)
\end{equation}
where $\gamma_{\omg}$ is now just the integrand of $\gamma_{\rm iso}$, i.e.
\begin{equation}\label{e:gammaOmegaRM}
\gamma_{\omg}=\frac{1}{2} \sum_A e^{i 2 \pi f \omg \cdot \frac{\dxvec}{c}}
    \,\, F_1^A(\omg) F_2^A(\omg)
\end{equation}
and $H(f)$ is the strain power spectrum of an unpolarized point source, summed over both polarizations.
Note that $\gamma_{\omg}$ also becomes sidereal time dependent both through $\dxvec$ and $F_i^A(\omg)$.

Eq. \ref{e:muPOmega} was used in Ref. \cite{allenottewill} as a starting point to derive optimal filters
for each spherical harmonic.
However if one wants to optimize the method for well located astrophysical sources it seems more
natural to use a $P(\omg)$ that only covers a localized patch in the sky. Indeed, for most
reasonable choices of $H(f)$, the resulting maximal resolution of this method will be bigger
than a several tens of square degrees such that most astrophysical sources won't be resolved.
Therefore it makes sense to optimize the method for true point
sources, i.e. $P(\omg)=\delta^2(\omg,\omg')$.

With this choice of $P(\omg)$ the optimal filter $Q_{\omg'}$ for the sky direction $\omg'$
becomes
\begin{equation}\label{e:optfiltRM}
Q_{\omg'}(f) = \lambda \frac{ \gamma_{\omg'}(f) H(f)}{P_1(f) P_2(f)}
\end{equation}
and the expected cross-correlation $Y_{\omg'}$ and its expected variance $\sigma^2_{Y_{\omg'}}$ are
\begin{equation}\label{e:sigmaRM}
    \sigma_{Y_{\omg'}}^2 \equiv \langle Y^2_{\omg'} \rangle - \langle Y_{\omg'} \rangle^2 \approx
               \frac{T}{4} \left( Q_{\omg'},Q_{\omg'} \right)
\: , \,\,\,\,\,\,\,\,\,\,\,\,
    \langle Y_{\omg'} \rangle = T \left( Q_{\omg'}, \frac{\gamma_{\omg'} H}{P_1 P_2}\right)
\end{equation}

\subsection{Integration over sidereal time}
Since the non-stationarity of the detector noise also needs to be dealt with, processing the
data on a segment by segment basis still makes sense. However $\gamma_{\omg}$ changes with
sidereal time. By setting it to its mid-segment value one can get rid of the 1st order error,
but a 2nd order error remains and is of the order
\begin{equation}\label{e:errorRM}
Y_{\rm err}(T_{\rm seg})/Y = 
\frac{{T_{\rm seg}}^2}{12} \frac{
\int_{-\infty}^{\infty} \frac{\partial^2\gamma_{\omg'}^*}{\partial t^2} \gamma_{\omg'}
\frac{H^2}{P_1 P_2} df
}{
\int_{-\infty}^{\infty} \gamma_{\omg'}^2 \frac{H^2}{P_1 P_2} df
} = O \left( \left( \frac{2 \pi f d}{c} \frac{T_{\rm seg}}{1~{\rm day}} \right)^2 \right)
\end{equation}
with $f$ the typical frequency and $d$ the detector separation. For $T_{\rm seg}=60~{\rm sec}$,
$f=2~{\rm kHz}$ and $d=3000~{\rm km}$ this error is less than 1\%.

Thus the integration over sidereal time for each $\omg'$ again reduces to the optimal combination of
the set $\{Y_I, \sigma_{Y_I}\}_{\omg'}$ given by Eq. \ref{e:Yopt}. The only difference
to the isotropic $P(\omg)\!=\!1$ case is that the optimal filter $Q_{\omg'}$ is different for each
interval $I$ and each sky direction $\omg'$.

\subsection{Numerical aspects}
To implement this method one thus has to calculate
\begin{equation}\label{e:intYRM}
Y_{\omg'} = \lambda T \int_{-\infty}^{\infty} df \frac{\gamma_{\omg'}^* H}{P_1 P_2} S1^* S2
\: , \,\,\,\,\,\,\,\,\,\,\,\,
\sigma_{\omg'}^2 = \lambda^2 \frac{T}{4} \int_{-\infty}^{\infty} df
 \frac{|\gamma_{\omg'}|^2 H^2}{P_1 P_2}
\end{equation}
for each sky direction $\omg'$ and each segment $I$. This can be calculated very
efficiently by realizing that $\gamma_{\omg}$ splits into a DC part
${1}/{2} \sum_A  F_1^A(\omg) F_2^A(\omg)$ and a phasor $\exp({i 2 \pi f \omg 
\cdot {\dxvec}/{c}})$. For both integrals
 the DC part can
be taken out of the frequency
integration, leaving all the directional information of the integrands in the phasor. Thus, with $N$
the number of sky directions $\omg'$, instead having to calculate $2 N$ integrations, it is
sufficient to calculate one fast Fourier transform and one integral per segment, and read out the
cross-correlation $Y_{\omg'}$ at the time shift $\tau = {\omg'} \cdot \frac{\dxvec}{c}$.

Since the fast Fourier transform of ${S1^* S2\,H} / ( {P_1 P_2} )$ is sampled at $f_{\rm sample} = 2f_{\rm Nyquist}$
it is necessary to interpolate to get the cross-correlation $Y_{\omg'}$ at the time shift $\tau$.
However, by choosing a high enough Nyquist frequency and zero-padding the unused bandwidth this
interpolation error can be kept small while the overall efficiency is still improved.

\subsection{Comparison to the isotropic case}
It is interesting to look at the potential signal-to-noise ratio improvement of this directional method
compared to the isotropic method if indeed all correlated signal would come from one point $\omg'$,
i.e. $\langle S_1^* S_2 \rangle = \gamma_{\omg'} H$.
The ratio between the two signal-to-noise ratios works out to
%
%
%
\begin{equation}\label{e:snrImpRM}
\frac{{\rm SNR}_{\rm iso}}{{\rm SNR}_{\omg'}} =
\frac{\langle Y_{\rm iso}^{\rm opt} \rangle  /\sigma_{\rm iso}^{\rm opt}}
{\langle Y_{\omg'}^{\rm opt} \rangle /\sigma_{\omg'}^{\rm opt}} =
\frac{\left[ \gamma_{\rm iso},\gamma_{\omg'} \right]}
{\sqrt{\left[ \gamma_{\rm iso},\gamma_{\rm iso} \right]\left[ \gamma_{\omg'},\gamma_{\omg'} \right]}}
\end{equation}
%
%
%
with $\left[ A,B \right] = \sum_i \left( A_i \HpPP,B_i \HpPP \right)$ and i the index summing over sidereal time.
This ratio is bounded between $-1$ and $1$, i.e. the directional search not only performs better
in this case but, for a point source at an unfortunate position, the isotropic search can even yield
negative or zero correlation.

It is also possible to recover the isotropic result as an integral over the sky. The definitions
of $\gamma_{\rm iso}$ and $\gamma_{\omg}$ (Eq. \ref{e:gamma_iso} and \ref{e:gammaOmegaRM}) imply
\begin{equation}\label{e:as1RM}
Y_{\rm iso}^{\rm opt} {\sigma_{\rm iso}^{\rm opt}}^{-2} = \frac{5}{4 \pi} \int d\omg \,\,\,
Y_{\omg}^{\rm opt}{\sigma_{\omg}^{\rm opt}}^{-2}
\: , \,\,\,\,\,\,\,\,\,\,\,\,
{\sigma_{\rm iso}^{\rm opt}}^{-2} = \left( {\frac{5}{4 \pi}} \right) ^2 
\int d\omg \int d\omg' {\sigma_{\omg,\omg'}^{\rm opt}}^{-2}
\end{equation}
Here $\sigma_{\omg,\omg'}^2$ is the covariance of the 2 sky directions $\omg$ and $\omg'$.
It is given by
\begin{equation}\label{e:sigmaRMcov}
    \sigma_{Y_{\omg,\omg'}}^2 \approx
               \frac{T}{4} \left( Q_{\omg},Q_{\omg'} \right)
\end{equation}
and describes the antenna lobe of the gravitational wave radiometer.

\subsection{Achievable sensitivity}
The 1-$\sigma$ sensitivity of this method is given by
\begin{equation}\label{e:sensitivity}
    H_{{\rm sens},\omg}(f) = \frac{\sigma_\omg}{T} H(f) = \frac{H(f)}{2 \sqrt{T} \sqrt{
    \langle \int_{-\infty}^{\infty} \frac{|\gamma_\omg|^2 H^2}{P_1 P_2} df \rangle_{\rm sidereal~day}
    }}
\end{equation}
$H_{\rm sens}$ is somewhat dependent on the declination and in theory
independent of right ascension. In practice though an uneven coverage
of the sidereal day due to downtime and time-of-day dependent sensitivity
will break this symmetry, leaving only an antipodal symmetry.

For the initial LIGO Hanford 4km - Livingston 4km pair (H1-L1), both at design sensitivity,
and a flat source power spectrum $H(f)={\rm const}$ this works out to
\begin{equation}\label{e:sensitivitySimple}
    H_{\rm sens}^{H1-L1} \approx 1.5 \times 10^{-50} ~{\rm Hz}^{-1} 
    \left( \frac{1~{\rm yr}}{T} \right)^{\frac{1}{2}} 
\end{equation}
with a 35\% variation depending on the declination. This corresponds to a gravitational wave
energy flux density
of
\begin{equation}\label{e:sensitivityJy}
F_{\rm gw} df \approx
    5 \times 10^{-11}~\frac{\rm Watt}{{\rm m}^2 ~{\rm Hz}}
    \cdot \left(\frac{f}{100~{\rm Hz}} \right)^2
    \cdot \left( \frac{1~{\rm yr}}{T} \right)^{\frac{1}{2}} df
\end{equation}

\subsection{Results from simulated data}
The described algorithm was implemented such that it is ready to run on real LIGO data. However, in order
to test the code, the real data was blanked out and simulated Gaussian noise uncorrelated between the 2 detectors and 
with a power spectrum shape
equal to the LIGO design sensitivity was added. Real lock segment start and stop times were used.
This takes into account the non-uniform day coverage. To get a quicker turn-around during testing the code was only
run on $1.7$ days of integrated simulated data. The signal power spectrum was assumed to be flat, $H(f)={\rm const}$.

\begin{figure}[htbp!]
\begin{center}
  \includegraphics[width=6in,angle=0]{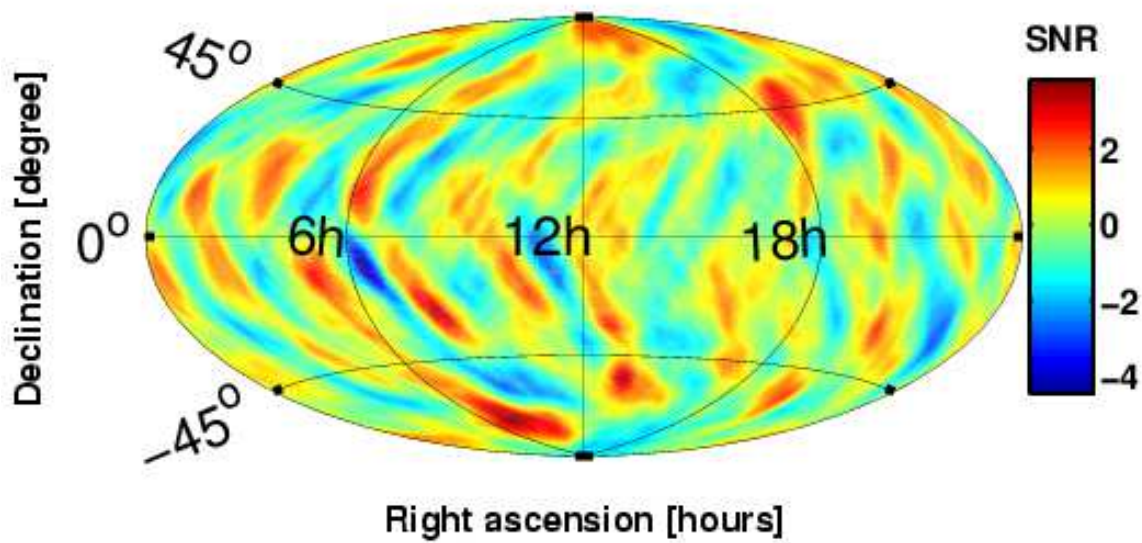}
  \caption{An example map of the signal-to-noise ratio ${\rm SNR}=Y_{\rm opt} / \sigma_{\rm opt}$ for simulated Gaussian noise (see text).
  The visible structure - fringes with opposite tilt on the northern and southern hemisphere as required by the antipodal symmetry of the antenna lobe -
  is due to the antenna lobe with which the random background is convolved. }
  \label{fig:snr}
\end{center}
\end{figure}

The algorithm was run on a 360 $\times$ 181 point grid covering the whole sky. While this clearly over-samples the intrinsic resolution -
for the $H(f)={\rm const}$ case the antenna lobe has a FWHM area of $50-100~{\rm deg}^2$, depending on the declination - it produces
nicer pictures as final product as shown in Figure \ref{fig:snr}.
%
%
%
%
\begin{figure}[htbp!]
\begin{center}
  \includegraphics[width=6in,angle=0]{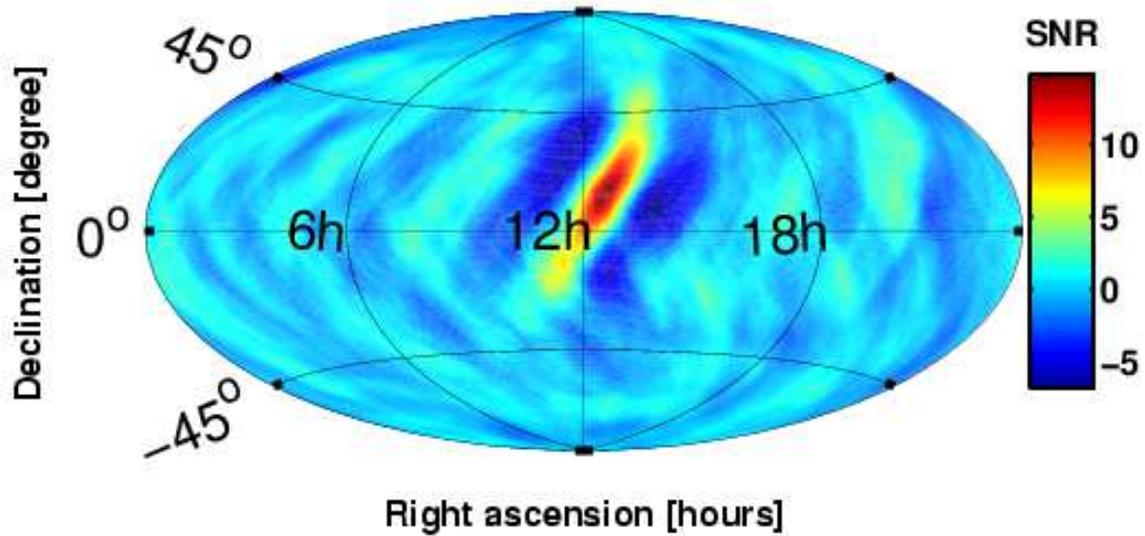}
  \caption{A point source with signal-to-noise ratio (SNR) of 14 was injected at the position of the Virgo galaxy cluster ($\rm12.5 h$, $12.7^\circ$). It is nicely recovered. This map also shows
  the typical structure of the antenna lobe including negative correlation in regions adjoint to the main peak.}
  \label{fig:snrI}
\end{center}
\end{figure}
%


For Figure \ref{fig:snrI} additional coherent noise with a flat power spectrum $H(f)$ and a sidereal time dependent time shift and amplitude modulation appropriate
for a true point source was added. The data was produced by splicing together short pieces of data with fixed time shift.
The injected source has a signal-to-noise ratio of 14 and is clearly recovered. This figure also shows the typical shape of the radiometer antenna lobe which is
given by Eq. \ref{e:sigmaRMcov}. In particular it shows that areas adjoint to the main lobe
get a negative correlation. This means that a particularly unfortunate set of sources could
in principle cancel a lot of the signal.

\section{Conclusion}
The presented gravitational wave radiometer method aims for optimal detection of
localized stochastic gravitational wave sources with a given power spectrum $H(f)$
and significant uptime. It produces a map of point estimates and corresponding variances.
The point estimate map corresponds to the true power distribution of gravitational waves $P(\omg)$ convolved
with the antenna lobe of the radiometer and an uncertainty determined by the detector noise. This antenna lobe in turn depends on the assumed
source power spectrum $H(f)$.

The method is well suited for searching for a stochastic gravitational wave background of astrophysical origin and
the LIGO Scientific Collaboration intends to use it on future LIGO science runs starting with S4.

The author gratefully acknowledges his colleagues in the LIGO Scientific Collaboration
whose work on building and commissioning ground based gravitational wave detectors
as well as computing infrastructure made this research possible.
In addition he acknowledges the support of the United States National 
Science Foundation for the construction and operation of 
LIGO under Cooperative Agreement PHY-0107417.


\end{document}